\documentclass[
preprint,
showpacs,
amsmath,
amssymb,
aps,
pra,
]{revtex4}

\usepackage{verbatim}
\usepackage{longtable}
\usepackage{multirow}
\usepackage{graphicx}
\usepackage{epstopdf}

\begin{document}



\newcommand{\ajp}{AJP}
\title{Associative detachment of \boldmath{${\rm H}^- + {\rm H} \to {\rm H}_2 + e^-$}}


\author{K. A. Miller,$^1$  
H. Bruhns,$^{1}$\footnote{Inficon GmbH, D-50968 Cologne, Germany.} 
J. Eli\'{a}\v{s}ek,$^2$  
M. \v{C}\'{\i}\v{z}ek,$^2$ 
H. Kreckel,$^1$\footnote{Department of Chemistry, University of Illinois, 600 South Mathews Avenue, Urbana, Il 61801, USA. \\ \copyright{American Physical Society 2011}
} 
X. Urbain,$^3$ and 
D. W. Savin$^1$}

\affiliation{
\begin{math}^{1}\end{math}Columbia Astrophysics Laboratory,Columbia University, New York, NY 10027, USA\\
\begin{math}^{2}\end{math}Charles University Prague, Faculty of Mathematics and Physics, Institute of Theoretical Physics, 180 00 Praha 8, Czech Republic\\
\begin{math}^{3}\end{math}Institute of Condensed Matter and Nanosciences, Universit\'{e} Catholique de Louvain, Louvain-la-Neuve B-1348, Belgium}


\date{\today}

\begin{abstract}

Using a merged beams apparatus we have measured the associative
detachment (AD) reaction of ${\rm H}^- + {\rm H} \to {\rm H}_2 + e^-$
for relative collision energies up to \textit{E}$_{\rm{r}} \leq
4.83$~eV.  These data extend above the 1~eV limit of our earlier
results.  We have also updated our previous theoretical work to
account for AD via the repulsive $^2\Sigma^+_{\rm g}$ H$_2^-$
potential energy surface and for the effects at $E_{\rm r} \ge
0.76$~eV on the experimental results due to the formation of H$_2$
resonances lying above the ${\rm H} + {\rm H}$ separated atoms limit.
Merging both experimental data sets, our results are in good agreement
with our new theoretical calculations and confirm the prediction that
this reaction essentially turns off for $E_{\rm r} \gtrsim 2$~eV.
Similar behavior has been predicted for the formation of protonium
from collisions of antiprotons and hydrogen atoms.

\end{abstract}

\pacs{34.50.Lf, 52.20.Hv, 95.30.Ft, 97.10.Bt}

\maketitle


\section{Introduction}

One of the simplest molecular formation reactions is associative
detachment (AD) via
\begin{equation} 
\mathrm{H^{-} + H 
\rightarrow  H_{2} + \textit{e}^{-}.} 
\label{eq:AD}
\end{equation}
This reaction is of interest for fundamental atomic and molecular
physics and also because it plays an important role in protogalactic
and first star formation in the early universe \cite{Glover1, Glover2,
  Holger}.  Two groups have recently reported measurements of this
reaction.  Martinez et al.~\cite{Martinez} measured the thermal rate
coefficient at 300~K using a flowing afterglow technique. Our group
has measured this reaction over a collision energy range from 4~meV to
1~eV using a merged-beams method \cite{Holger, Bruhns1, Bruhns2}.  Our
results lie $2.2 \pm 0.9$ times above those of \cite{Martinez}.  The
quoted uncertainty represents the quadrature sum of the estimated
total experimental 1$\sigma$ confidence level for each measurement;
and we have also taken into account minor corrections to our earlier
data which are described below.

In \cite{Bruhns2} we hypothesized that this discrepancy is due to an
error in the measured rate coefficient of \cite{Howard} for
\begin{equation}
\mathrm{H} + \mathrm{Cl^-} \rightarrow \mathrm{HCl} + e^-
\label{eq:HCl}
\end{equation}
which \cite{Martinez} used to determine their neutral H number density
and thereby normalize their results.  Our apparatus is not configured
to study reaction~(\ref{eq:HCl}) and test this hypothesis, but we have
been able to extend our measurements of reaction~(\ref{eq:AD}) to
higher energies and thereby provide additional benchmarks for theory.
We have also investigated and ruled out several possible sources of
systematic errors in our previous experimental results.  Additionally
we have more carefully considered the pressure dependence of our
detection method.  Lastly, we have updated our previous theoretical
results of \cite{Holger, Cizek} to account for AD via the repulsive
$^2\Sigma_{\rm g}^+$ H$_2^-$ state and for the effects at $E_{\rm r}
\ge 0.76$~eV on the experimental results due to the formation of
long-lived H$_2$ resonances lying above the ${\rm H} + {\rm H}$
separated atoms limit.

The rest of the paper is organized as follows:
Section~\ref{sec:Experiment} describes the experimental method and the
various modifications performed for this work.
Section~\ref{sec:Uncertainties} discusses the experimental
uncertainties.  Our new theoretical calculations are briefly described
in Sec.~\ref{sec:Theory}.  In Sec.~\ref{sec:Results} we present
our results and compare them to theory. A discussion of our results is
given in Sec.~\ref{sec:Discussion} and a short summary in
Sec.~\ref{sec:Summary}.


\section{Experiment}
\label{sec:Experiment}

Here we briefly describe the experiment and the
changes relevant to our new results.  Further details about
the apparatus and experimental method can be found in 
\cite{Holger, Bruhns1, Bruhns2}.


\subsection{Method}

We begin by extracting H$^{-}$ from a duoplasmatron source and forming
a beam with an energy of $E_{\rm H^-}=-e(U_{\rm s} + U_{\rm f}/2)$.
Here $e$ is the unit charge, $U_{\rm s} \approx - 10$~kV is the nominal
source voltage, and $U_{\rm f}$ is a small correction voltage defined
below.  Using standard ion optical elements, we shape, steer, and
direct the beam into a photodetachment chamber which houses a floating
cell biased to a potential $U_{\rm f}$.  The anion energy inside the
floating cell is $E_{\rm H^-} = -e(U_{\rm s} + U_{\rm f}/2) + eU_{\rm
  f}$.  Near the center of the floating cell, we cross the anions with
an infrared laser and convert a portion of the H$^-$ beam into a beam
of ground state H atoms of energy $E_{\rm H} = -e(U_{\rm s} - U_{\rm
  f}/2)$.  The resulting merged beams exit the floating cell,
whereupon the H$^-$ beam returns to its initial energy while the H
beam energy remains unchanged.  The beam-beam interaction energy is
controlled by varying $U_{\rm f}$.
  
Shortly after leaving the photodetachment chamber, the two beams enter
an interaction region of length $L$.  Two beam profile monitors (BPMs)
are used to determine the beam-beam overlap $\langle \Omega(z)
\rangle$ within the interaction region, where the $z$ axis is defined
by the bulk velocity vectors of the co-propagating beams.  We also use
the BPMs to verify the alignment of the beam axes.  The relative
energy $E_{\mathrm{r}}$ between the beams depends, in part, on this
alignment and is given by \cite{Phaneuf} as
\begin{equation}
E_{\mathrm{r}} = 
\mu 
\left( 
\frac{E_{\rm H^-}}{m_{\rm H^-}} + 
\frac{E_{\rm H}}{m_{\rm H}} - 
2\sqrt{\frac{E_{\rm H^-} E_{\rm H}}{m_{\rm H^-} m_{\rm H}}} \cos\theta
\right).
\label{eq:Er}
\end{equation}
Here $\mu = m_{\rm H^-} m_{\rm H}/(m_{\rm H^-} + m_{\rm H})$ is the
reduced mass of the colliding system; $m_{\rm H^-}$ and $m_{\rm H}$
are the masses of the H$^-$ and H, respectively; and $\theta$ is the
angle of intersection.  $E_{\rm r}$ is controlled by varying $U_{\rm
  f}$.  This merged beams approach allows us to reach collision
energies on the order of a few meV, limited only by the alignment of
the beams, the spread in collision angles between the two beams, and
the energy spread of each beam. We used geometrical simulations
\cite{Holger, Bruhns1, Bruhns2} to determine the average collision
energy $\langle E_{\rm r} \rangle$ versus $U_{\rm f}$, taking into
account the spreads in beam energies and angles.

Both beams are chopped out of phase in order to extract the signal
H$_2$ generated in the interaction region from various backgrounds.
Any H$_2$ formed in the interaction region has an energy of $E_{\rm
  H_2} = E_{\rm H^-} + E_{\rm H} = -2eU_{\rm s} = 20$~keV, neglecting
the $\lesssim 3.7$~eV kinetic energy of the detached electron.  At the
end of this region, an electrostatic quadrupole deflector is used to
direct the H$^-$ into a Faraday cup where the current \textit{I}$_{\rm
  H^-}$ is read and recorded.  The parent H and daughter H$_2$ beams
continue on into a gas cell kept at a helium pressure of $2 \times
10^{-4}$~Torr for most measurements.  Inside the cell a fraction of
the H$_2$ is ionized by the stripping collisions forming $\approx
20$~keV H$_2^+$.  Additionally, stripping of the H beam and
dissociative ionization of the H$_2$ can produce $\approx 10$~keV
H$^+$.

After the gas cell, the neutrals and resulting ions enter the analyzer
region involving two double-focusing, electrostatic cylindrical
deflectors in series \cite{Holger2} and a channel electron multiplier
(CEM).  A hole in the outer plate of the first or lower cylindrical
deflector (LCD) allows neutrals to pass through and travel into a
neutral detector.  The neutral particle current $I_{\rm H}$, as
measured in amperes, is monitored by measuring the secondary negative
particle emission from the target inside the neutral detector.  The
voltages on the LCD and upper cylindrical deflector (UCD) are selected
to transmit the 20~keV H$_2^+$ signal ions into the CEM while
rejecting any of the 10~keV H$^+$ formed in the gas cell.  

We study reaction~(\ref{eq:AD}) from the number of H$_2^+$ ions
detected in the CEM. Experimentally, we measure the cross section
$\sigma_{\rm AD}$ times the relative velocity $v_{\rm r}$ between the
H$^{-}$ and H beams convolved with the velocity spread of the
experiment.  This gives the rate coefficient \cite{Bruhns2}
\begin{equation} 
\langle \sigma_{\rm AD}v_{\rm r} \rangle = 
\frac{1}{\sigma_{\rm st}N_{\rm He}} 
\frac{S}{T_{\rm a}T_{\rm g}\eta}
\frac{e^2}{I_{\rm H^-}I_{\rm H}}
\frac{v_{\rm H^-}v_{\rm H}}{L\langle \Omega(z) \rangle}. 
\label{eq:rate}
\end{equation} 
The left hand side average is over the experimental energy spread.  
On the right side, $\sigma_{\rm st}$ is the
stripping cross section for H$_2$ on He forming H$_2^+$; $N_{\rm He}$
is the gas cell helium column density; $S$ is the background
subtracted, pressure corrected H$_2^+$ signal; $T_{\rm a}$ is the
transmittance of the combined LCD-UCD analyzer; $T_{\rm g}$ is the
transmittance of the grid in front of the CEM; $\eta$ is the CEM
efficiency; and $v_{\rm H^-}$ and $v_{\rm H}$ are the velocities of
the H$^-$ and H beams, respectively.


\subsection{Modifications}

The present work uses a current meter with a fast response time which
enables us to directly measure the H$^-$ current at each phase in the
chopping pattern, which is on the millisecond scale, and monitor it
throughout each data run.  Thus we are able to measure the anion
current when the laser is on, $I_{\rm H^-}^{\rm on}$, over the course
of a data run. This is used for $I_{\rm H^-}$ in Eq.~(\ref{eq:rate}).
We were also able to monitor the anion current with the laser off,
$I_{\rm H^-}^{\rm off}$, and determine the attenuation factor
\begin{equation}
f = 1 - \frac{I_{\rm H^{-}}^{\rm on}}{I_{\rm H^{-}}^{\rm off}},
\label{eq:f}
\end{equation}
which is needed to extract the background corrected $S$
\cite{Bruhns2}.  This situation is to be contrasted with our previous
results \cite{Holger, Bruhns2} where, due to equipment limitations,
the H$^-$ current was averaged over the H$^-$ chopping cycle and the
resulting $\left\langle I_{\rm H^-}^{\rm chop}\right\rangle$ was
recorded using a slow current meter. As a result, for that work $f$
was not measured during data collection but under simulated data
collection conditions and an average value was used. Additionally this
factor was used to extract $I_{\rm H^-}^{\rm on}$ and $I_{\rm
  H^-}^{\rm off}$ from $\langle I_{\rm H^-}^{\rm chop}\rangle$.

For the present work we are also using a new calibrated neutral
detector in combination with a fast current amplifier to record the H
particle current at each phase in the chopping pattern and to monitor
it throughout each data run.  This modification is described in
\cite{Bruhns2}.  Thus, during a data run, we are now able to directly
measure $I_{\rm H}$ which is needed in Eq.~(\ref{eq:rate}). In our
previous work, the H particle current was also monitored with a fast
current amplifier; however, the neutral detector was not designed for
absolute measurements.  So to analyze those results, using the new
detector we measured the H particle current due to photodetachment
(PD), $I_{\rm H}^{\rm PD}$, to determine the neutral-to-anion (nta)
ratio
\begin{equation}
f_{\rm nta} = \frac{I^{\rm PD}_{\rm H}}{I_{\rm H^-}^{\rm off}}
\label{eq:nta}
\end{equation}
under simulated data collection conditions.  This factor, combined
with the extracted $I_{\rm H^-}^{\rm off}$ discussed above, was then
used in \cite{Bruhns2} to determine $I_{\rm H}$ for
Eq.~(\ref{eq:rate}).

We have also installed a BPM immediately before the neutral detector,
at a distance of 2055~mm from the first BPM in the interaction region.
Turning off the voltage of the LCD, allows the H$^-$ beam to pass
through the hole in the outer plate of the LCD.  We used this
additional BPM to measure the position of both the H and H$^-$ beams
and verified the alignment of the beams over a much longer lever arm
than was previously possible using only the two BPMs in the
interaction region. We find that the full angle between the beam axes
measured here is in good agreement with that reported in
\cite{Bruhns2}.


\subsection{Pressure corrections}
\label{subsec:PressureCorrections}

Any H$_2^+$ formed in the gas cell can be destroyed by subsequent
collisions with He in either the gas cell or the analyzer region.  The
resulting products are not transmitted by the electrostatic deflectors
into the CEM, thereby reducing the apparent signal and rate
coefficient.  This small systematic shift to our data was
overlooked in our previous work \cite{Bruhns2, Holger}.  Here we
quantified this minor correction for both our previous and present
results.

We measured the H$_2^+$ attenuation using an approach similar to the
one we used to determine the He gas cell column density in
\cite{Bruhns2, Holger}. Reconfiguring the ion source to produce
H$_2^+$ and the apparatus to transmit H$_2^+$ beams, we used the
electrostatic quadrupole after the interaction region to direct the
beam into a Faraday cup where we measured the unattenuated H$_2^+$
current, $I^{\rm o}_{\rm H_{2}^{+}}$.  We then guided the beam through
the gas cell and measured the transmitted current, $I_{\rm
  H_{2}^{+}}$, on the outer plate of the UCD. With no He in the gas
cell, the UCD reading was over 95\% of that in the Faraday cup.  The
measured attenuated data were corrected for this slight difference in
the unattenuated current readings.

The H$_{2}^{+}$ attenuation as a function of gas density is given by
\begin{equation} 
\frac{I_{\rm H_{2}^{+}}}{I^{\rm o}_{\rm H_{2}^{+}}} = 
\rm{exp}(-\sigma_{\rm d}\textsl{N}_{\rm He}), 
\label{eq:H2atten} 
\end{equation}
where $\sigma_{\rm d}$ is the total H$_{2}^{+}$ destruction cross
section and $N_{\rm He}$ is the helium column density.  Following the
methodology of \cite{Bruhns2}, the column density can be expressed as
\begin{equation}
\label{eq:NHeint}
N_{\rm He} = 
\int_{\rm quad}{n_{\rm He}(l)dl} + 
\int_{\textrm{gas cell}}{n_{\rm He}(l)dl} + 
\int_{\rm analyzer}{n_{\rm He}(l)dl}.
\end{equation}
Here $n_{\rm He}(l)$ is the helium density and $dl$ the
infinitesimal path length.  Using the same model as \cite{Bruhns2}, we
take the pressure to be constant in each of these regions and
re-express Eq.~\ref{eq:NHeint} as
\begin{equation}
\label{eq:NHesimple}
N_{\rm He} = n_{1}l_{1} + n_{2}l_{2} + n_{3}l_{3}.  
\end{equation}
The He density in the quadrupole is $n_1$ and the path length $l_1 =
5.0 \pm 1.0$~cm.  In the gas cell the He density is $n_2$ and the path
length $l_2 = 78.7 \pm 1.0$~cm.  The He density in the analyzer region
is $n_3$ and the path length $l_3 = 35.4 \pm 1.0$~cm is the distance
that the ions travel before striking the UCD.  All uncertainties here
and throughout the paper are given at an estimated 1$\sigma$
statistical confidence level.  The respective densities were
calculated from the measured pressures using the ideal gas law at the
laboratory temperature which was stabilized at 293~K for both the work
of \cite{Holger, Bruhns2} and our new results here.  The ratio of the
measured pressures in each section were $p_1/p_2 = 0.137 \pm 0.019$
and $p_3/p_2 = 0.105 \pm 0.034$.  The uncertainties in these ratios
are due to the manufacturer-quoted accuracies of the pressure gauges
(10\% for $p_1$ and $p_2$ and 30\% for $p_3$).

\begin{figure}[!h]
\includegraphics[width= 1\textwidth]{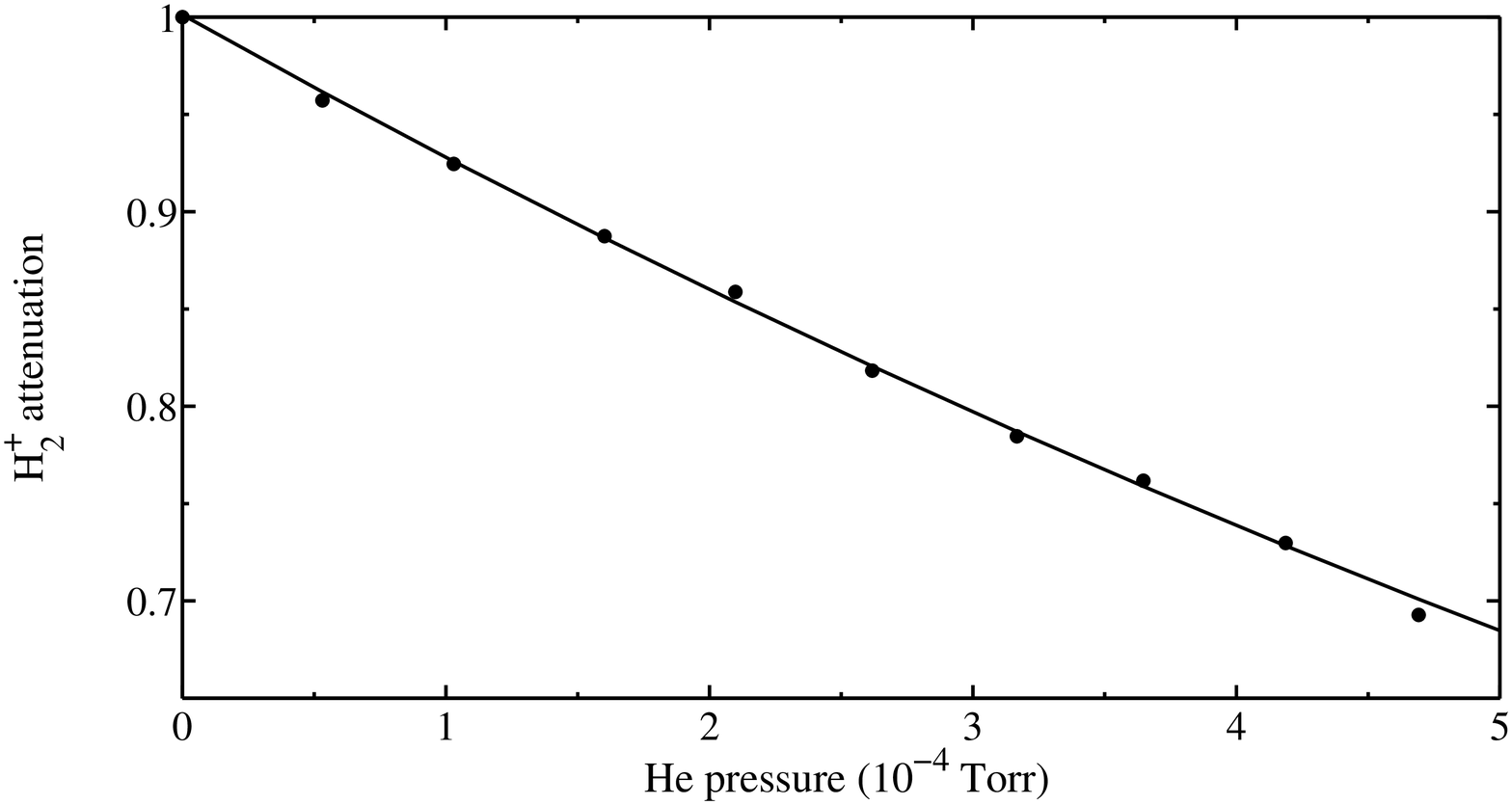}
\caption{Attenuation of the H$_2^+$ ion beam as a function of helium
  gas cell pressure.  The circles represent the statistically-weighted
  mean from three sets of measurements.  The error bars are smaller
  than the plotted circles.  The line shows the best exponential fit.}
\label{fig:Attenuation} 
\end{figure}

Attenuation data were collected for pressures up to $\approx 4.5 \times
10^{-4}$~Torr and are shown in Fig.~\ref{fig:Attenuation}.  From a fit
to these data we extracted a cross section of $(2.75 \pm
0.29)\times10^{-16}$ cm$^{2}$ at an energy of 10~keV~amu$^{-1}$.  This
estimated uncertainty is due to the error in the attenuated and
unattenuated current readings (3\% each) and the uncertainty in the He
column density~(10\%).  The error in this latter quantity was
estimated by adding the uncertainties from each segment $n_{i}l_{i}$
of the total column density. The errors in the path lengths and gas
densities (i.e., pressures) have been given above.

\begin{figure}[!h] 
\includegraphics[width= 1\textwidth]{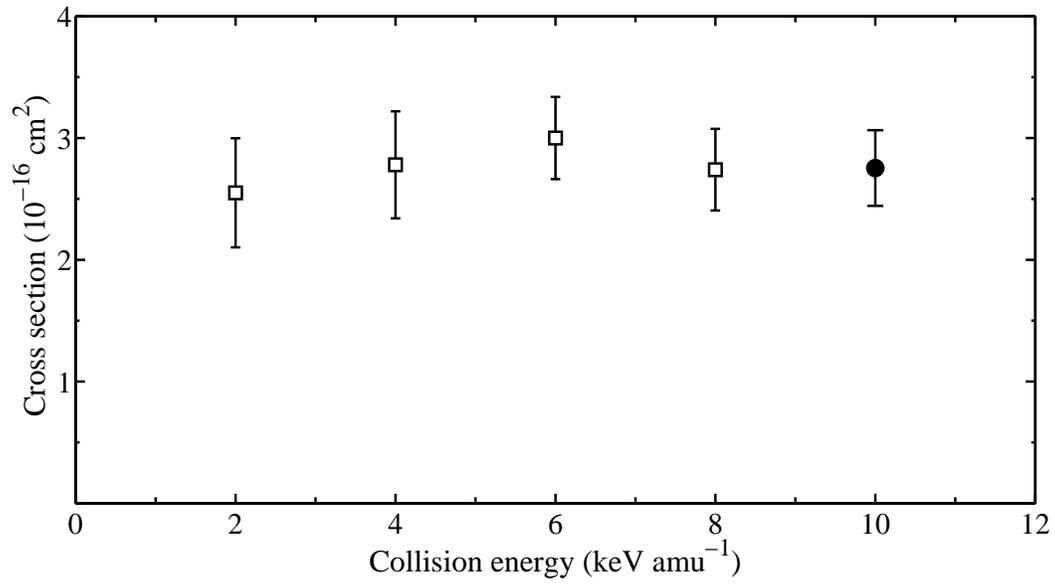}
\caption{Experimental cross sections for total H$_{2}^{+}$ destruction
  versus ion beam energy for ${\rm H}_{2}^{+} + {\rm He}$.  The open
  squares are the results of \cite{Suzuki} while the circle
  represents our measurement.  The error bars for each data set give
  the total 1$\sigma$ experimental uncertainty.}
\label{fig:Suzuki} 
\end{figure}

Collisional destruction of H$_2^+$ has also been studied by
\cite{Suzuki} who reported cross sections of various outgoing channels
for ion energies from 2 - 8 keV amu$^{-1}$.  We have derived a total
destruction cross section by summing the relevant channels in
\cite{Suzuki}.  Those results, shown in Fig.~\ref{fig:Suzuki},
indicate that the cross section is essentially constant between
2~and~8~keV~amu$^{-1}$.  Our result at 10~keV~amu$^{-1}$, also shown
in Fig.~\ref{fig:Suzuki}, is in good agreement with this trend.

To determine the expected signal attenuation factor and correct for
the H$_2^+$ signal loss we use our measured H$_2^+$ destruction cross
section combined with Eq.~(\ref{eq:H2atten}).  The appropriate He
column density is given by
\begin{equation}
N^\prime_{\rm He} = 
\frac{1}{2}n_{2}l_{2} + n_{3}l'_{3}, 
\label{eq:nhestrip} 
\end{equation}
where the factor of $1/2$ takes into account that on average the
H$_2^+$ ions will be formed in the center of the gas cell and
$l^\prime_3 = 57.9 \pm 1.0$~cm is the distance from the end of the gas
cell to the CEM mouth. Using these values we calculate from
Eq.~\ref{eq:H2atten} that the signal attenuation with $2 \times
10^{-4}$~Torr He in the gas cell is $0.92 \pm 0.01$.  The signal must
be divided by this factor to correct for the attenuation.  This
corresponds to an ($8.6 \pm 1.2$)\% upward shift in the data.  The
uncertainty in this correction is estimated by propagating through
Eq.~\ref{eq:H2atten} the quadrature sum of the uncertainties from both
$N_{\rm He}^\prime$ in Eq.~(\ref{eq:nhestrip}) and $\sigma_{d}$.


\section{Uncertainties}
\label{sec:Uncertainties}

The various systematic uncertainties for the measurement are given in
Table~\ref{tab:errors}.  Values are listed at an estimated $1\sigma$
statistical confidence level.  We have grouped them into two sets.
The errors listed in the top half of the table add in quadrature to
$\pm 12$\% for each data point.  This represents the relative
uncertainty between our old and new data sets and also at different
energies within each set.  Adding this in quadrature with the
remaining uncertainties in the bottom half of the table yields the
total systematic error of $\pm 24\%$.  A detailed discussion is given
in \cite{Holger, Bruhns2} for the various uncertainties not already
discussed here.


\section{Theory}
\label{sec:Theory}

\subsection{Earlier calculations}

\begin{figure}[!h]
\includegraphics[width= 1\textwidth]{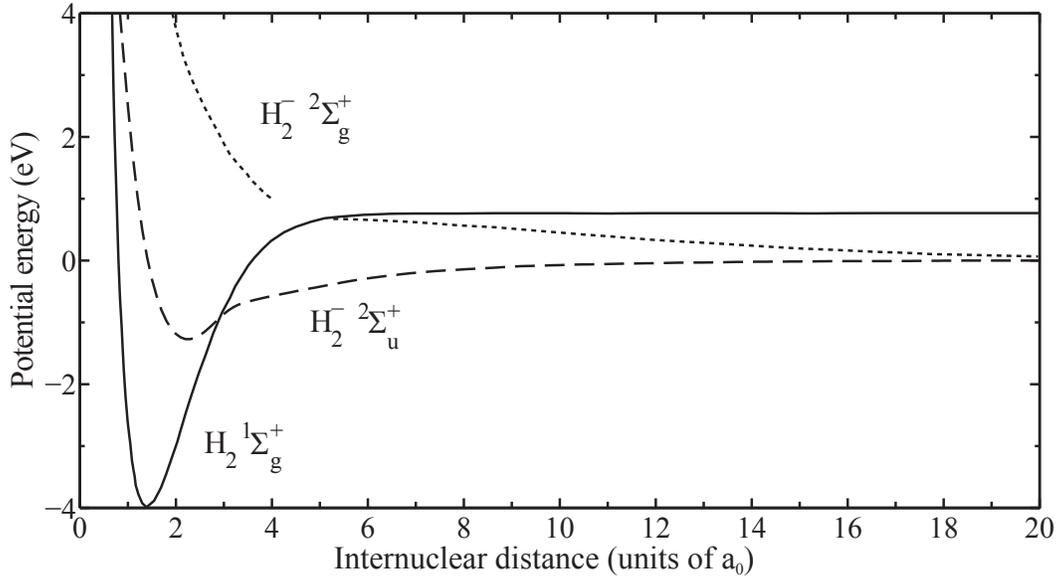}
\caption{H$_{2}^{-}$ and H$_{2}$ potential curves versus internuclear
  distance in units of the Bohr radius $a_0$. The H$_{2}^{-}$
  attractive $^{2}\Sigma^{+}_{\rm u}$ electronic state is given by the
  dashed curve \cite{Senekowitsch} and the repulsive
  $^{2}\Sigma^{+}_{\rm g}$ electronic state by the dotted curve
  constructed using the data of \cite{Stibbe1, Stibbe2} below $\sim
  4~a_0$ and those of \cite{Bardsley} above $\sim 5~a_0$.  The
  separated atoms limit (SAL) for these two potential energy curves is
  ${\rm H}^{-}(^{1}{\rm S}) + {\rm H}(^{2}{\rm S})$.  The solid curve
  shows the H$_{2}$ $^{1}\Sigma^{+}_{\rm g}$ electronic state from
  \cite{Wolniewicz} with an SAL of ${\rm H}(^{2}{\rm S}) + {\rm
    H}(^{2}{\rm S}$).  The energy difference between the two limits is
  determined by the H electron affinity energy $E_{\rm EA} =
  0.76$~eV~\cite{Horacek}.}
\label{fig:pec}
\end{figure}

In our previous work, the AD cross section was calculated using
non-local resonance theory and considering only the coupling of the
${\rm H} + {\rm H}^-$ and ${\rm H}_2+e^-$ channels through the lowest
metastable H$_2^-$ state of $^2\Sigma^+_{\rm u}$ symmetry (see
\cite{Holger, Cizek} for details).  This state is one of two connected
to the ${\rm H} + {\rm H}^-$ asymptote (not counting the spin
degeneracy).  Potential energy curves for both states are shown in
Fig.~\ref{fig:pec}.  The second state of the $^2\Sigma^+_{\rm g}$
symmetry is repulsive and usually neglected in the calculations.  The
validity of this approximation is supported by the very good agreement
between our experimental results \cite{Holger,Bruhns2} and our
non-local calculations \cite{Holger,Cizek} below 1~eV, even after the
$\sim 9\%$ pressure correction of the H$_2^+$ signal described in
Sec.~\ref{subsec:PressureCorrections} which was not accounted for in
\cite{Holger,Bruhns2}.

\subsection{New calculations}

We have extended our experimental results to $\sim 5$~eV, entering a
regime where AD via the $^2\Sigma^+_{\rm g}$ state becomes possible.
Figure~\ref{fig:pec} shows that for sufficiently large energies the
colliding ${\rm H} + {\rm H}^-$ can penetrate into the autodetachment
region along the repulsive $^2\Sigma^+_{\rm g}$ state.  This region is
defined as the range of internuclear separations $R$ where an electron
can escape the anionic system, i.e., the potential energy curves of
the H$_2^-$ system are above those for neutral H$_2$.  This occurs for
the $^2\Sigma^+_{\rm g}$ state at $R<5~a_0$, where $a_0$ is the Bohr
radius.  Particles colliding along this state can penetrate into the
autodetachment region for energies $\gtrsim 0.75$~eV motivating
calculations for AD via this state.

Due to the different symmetry of the molecular orbitals, the
$^2\Sigma^+_{\rm u}$ and $^2\Sigma^+_{\rm g}$ contributions to the AD
cross section can be calculated separately.  Thus we need only carry
out new calculations for the $^2\Sigma^+_{\rm g}$ state.  A brief
description of our approach is presented below, using atomic units.  A
more detailed discussion will be given in a future publication.

Non-local resonance theory is explained in detail by \cite{d91}.  The
main idea is as follows.  The electronic state $\phi_d$, describing
the colliding partners in the ${\rm H} + {\rm H}^-$ channel, is
diabatically prolonged to small $R$.  It is also assumed to be coupled
to the ${\rm H}_2+e^-$ electronic continum states $\phi_k$ through the
matrix element
\begin{equation}
\nonumber
V_{dk}(R)=\langle\phi_d | H_{\rm el} | \phi_k\rangle
\end{equation}
where $H_{\rm el}$ is the electronic Hamiltonian.  The non-local
resonance model is parametrized by three functions: $V_0(R)$,
$V_d(R)$, and $V_{dk}(R)$.  The potential energy curve for the neutral
molecule $V_0(R)$ and for the anion $V_d(R)$, are functions only of
$R$.  The coupling element $V_{dk}(R)$, however, depends on both $R$
and the momentum of the detached electron $k$.

Once $V_0(R)$, $V_d(R)$, and $V_{dk}(R)$ are known, the electronic
dynamics of the system is fully parametrized and the nuclear dynamics
can be treated as a motion in the non-local energy-dependent effective
potential
\begin{equation}
\nonumber
V_d(R) + 
\int V_{dk} (E-{\textstyle\frac{1}{2}}k^2-T_N-V_0(R)+i\varepsilon)^{-1}V_{dk}^* k
\, dk\, d\Omega_k.
\end{equation}
$T_N$ is the kinetic energy operator for the nuclei, ${\rm d}\Omega_k$
is the differential solid angle for the outgoing electron, and
$\varepsilon$ is the usual positive infinitesimal of scattering
theory.  We solve the nuclear dynamics and calculate the cross
sections using the method of \cite{Cizek}.

In order to include $\phi_d$ for the $^2\Sigma^+_{\rm g}$ state, we
have to fix the parameters of the non-local resonance model for this
state.  The proper procedure for calculating these parameters involves
extracting the discrete state $\phi_d$ from the continuum $\phi_k$,
employing the projection-operator technique.  This procedure was
followed in \cite{mbd85} for the $^2\Sigma^+_{\rm u}$ state and we
used it as an input for our calculation \cite{Cizek}.  But it is also
possible to fix the model parameters by fitting the fixed nuclei
scattering data.  We follow this latter procedure here.

To fix the coupling amplitude, we assume the separable form
$V_{dk}(R)=g(R)f(k)$, where the $k$ dependence is determined by
the Wigner threshold law \cite{Wigner} with an exponential cut-off
\begin{equation}
\label{eq:wigner}
f(k)\sim k^{2l+1}e^{-\alpha k^2},
\end{equation}
where $\alpha$ is the cut-off parameter.  The angular momentum $l$
value in Eq.~(\ref{eq:wigner}) is given by the lowest electron partial
wave allowed by symmetry (discussed below).  The $R$ dependence is
determined from the calculated local decay widths $\Gamma(R)$ of
\cite{Stibbe1}.  The potential energy curve for the anion $V_d(R)$ is
constructed from \cite{Stibbe1,Stibbe2} and extended to larger $R$
using the data of \cite{Bardsley}.  The data for the potential near
the crossing of the neutral and anion potential energy curves are
missing.  Nevertheless the analytic behavior near the crossing has
been discussed in detail by \cite{d91}.  With this knowledge, the
potential energy curve can be interpolated through the crossing as has
been done before for hydrogen halides \cite{hx}.  The actual shape of
the $V_d(R)$ and $V_0(R)$ crossing is modified by the interaction of
the electron scattering continuum with the threshold behavior given by
the Eq.~(\ref{eq:wigner}).  $V_0$ is from \cite{Wolniewicz}.

The decay of the odd symmetry anion $^2\Sigma^+_{\rm u}$ state to the
even neutral $^1\Sigma^+_{\rm g}$ state is possible only through
release of an electron with odd angular momentum.  In
\cite{Holger,Cizek} we considered only $l=1$ ($p$-wave scattering)
since the calculations of \cite{Berm85a} show that the next allowed
$l=3$ contribution is suppressed by almost two orders of magnitude for
the energy range of interest.  For the anion $^2\Sigma^+_{\rm g}$
state decaying to the neutral $^1\Sigma^+_{\rm g}$ state, the symmetry
remains unchanged, requiring release of an electron with even
angular momentum.  Here we considered only $l=0$ ($s$-wave
scattering).  In each case, as Eq.~\ref{eq:wigner} shows, $V_{dk}$ is
strongly suppressed for higher angular momenta at the $k <
1$ values in our experimental results.

The anion $^2\Sigma^+_{\rm g}$ state can also decay to the first
excited $^3\Sigma^+_{\rm u}$ state of H$_2$.  These states are both
repulsive and lie very close together.  This decay, however, requires
an odd value for $l$.  With $l = 1$ and $k < 1$, $f(k)$, and hence
$V_{dk}$, is strongly suppressed compared to the $^2\Sigma^+_{\rm g}$
to $^1\Sigma^+_{\rm g}$ decay channel with $l=0$.  The effect from the
transition between repulsive states is thus expected to be small at
low energies and was not included here.

\begin{figure}[!h]
\includegraphics[width= 1\textwidth]{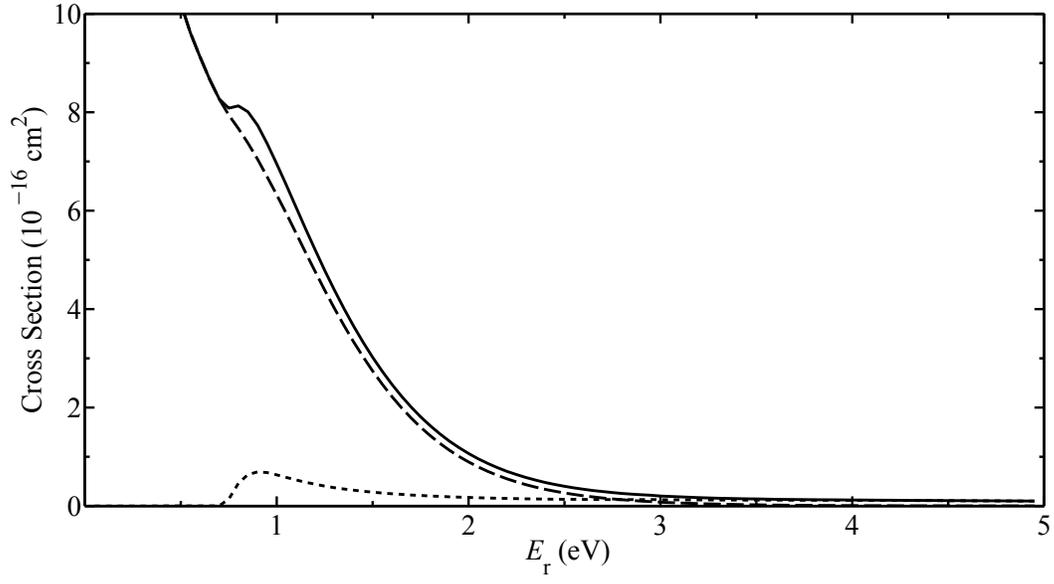}
\caption{Theoretical cross section for ${\rm H}^- + {\rm H} \to
  {\rm H}_2 + e^-$ as a function of the relative collision energy
  $E_{\rm r}$.  The dashed curve shows the results via the attractive
  H$_2^- \ ^2\Sigma^+_{\rm u}$ state, the dotted curve via the
  repulsive H$_2^- \ ^2\Sigma^+_{\rm g}$ state, and the solid curve
  the sum of the two.}
\label{fig:sAD}
\end{figure}

Once the model parameters are fixed, $\sigma_{\rm AD}$ can be
calculated using the methods described in \cite{Cizek}.
Figure~\ref{fig:sAD} shows our results.  As expected from the previous
good agreement of our experimental and theoretical results, the new
contribution is small and notable only for $E_{\rm r} \gtrsim
0.75$~eV.  This is the threshold where the colliding particles
overcome the barrier in the repulsive interaction potential and
penetrate into the autodetachment region.

Both the $^2\Sigma^+_{\rm u}$ and $^2\Sigma^+_{\rm g}$ contributions
decay rapidly to zero for energies above $\sim 1$~eV.  This is due to
the competing process of collisional detachment
\begin{equation}
\label{eq:CD}
{\rm H}^- + {\rm H} \to {\rm H} + {\rm H} + e^-,
\end{equation}
which opens up for $E_{\rm r} = 0.76$~eV and wins at higher energies.
This is discussed in Sec.~\ref{sec:Discussion} from the point of view
of general energy conservation arguments.

\subsection{Contributions from quasi-bound H$_2$ states}

At high angular momentum ($J > 10$), the colliding H$^-$ and H systems
can autodetach into quasi-bound H$_2$.  These states, sometimes
referred to as orbiting or shape resonances, lie above the separated
atoms limit for ${\rm H} + {\rm H}$.  Such high $J$ levels,
temporarily stabilized by the centrifugal barrier, will eventually
dissociate spontaneously and are therefore generally not considered in
AD cross section calculations.  However, the lifetime for all but a
few of these resonances well exceeds the flight time from the
interaction region to the gas cell and so most are expected to
contribute to the experimental signal.

The H$_2$ flight time from the interaction region to the gas cell is
$(737 \pm 640)$~ns.  The mean is the center-to-center distance, the
upper limit is from the start of the interaction region to the end of
the gas cell, and the lower limit from the end of the interaction
region to the start of the gas cell.  Quasi-bound H$_2$ (i.e., in high
$J$ levels) can strip in the He gas cell and will form H$_2^+$ in
similarly high $J$ levels.  As the H$_2^+$ potential supports stable
ro-vibrational levels up to $J = 35$, we assume that any such H$_2^+$
formed will be stable and will reach the detector.

In order to compare to our measured results, we have added the
contribution of these quasi-bound H$_2$ states to our calculations for
AD via the $^2\Sigma^+_{\rm u}$ state.  So as to mimic the range of
experimental lifetimes, we have investigated the effect of cutting out
states with lifetimes less than 100, 700, and 1400~ns and found no
significant differences.  In the end we included contributions from
all resonances with lifetimes longer than 700~ns.  The contribution of
these states is of comparable size to the $^2\Sigma^+_{\rm g}$ state
contribution. The effect of these resonances for AD via the
$^2\Sigma^+_{\rm g}$ state has not been considered as that would be a
small correction to an already small contribution.  Lastly, we note
that the significance of these resonances for molecular hydrogen
formation in plasma environments will depend on whether the states can
relax to stable states of H$_2$ before they dissociate by tunnelling.


\section{Results}
\label{sec:Results}


Relative energies \begin{math}E_{\rm{r}}\end{math} are controlled by
varying the potential of the floating cell $U_{\rm{f}}$.  In
\cite{Holger, Bruhns2} data were collected for $E_{\rm{r}} \leq 1
\rm{eV}$ $(|U_{\rm{f}}| \leq 281$ V).  Here we have extended the
energy range to $E_{\rm{r}} \leq 4.83$ eV $(|\textit{U}_{\rm{f}}| \leq
621$ V).  Data are collected by stepping $U_{\rm{f}}$ in voltage.  The
present work uses voltage ranges smaller than our earlier
measurements.  For $|U_{\rm{f}}| \leq 441$ V, $U_{\rm{f}}$ was scanned
across 60 V ranges in 10 V steps and for $|{U_{\rm{f}}|} \geq 441$ V
the scanning was across 120~V ranges in 20~V steps.

\begin{figure}[!h]
\includegraphics[width= 1\textwidth]{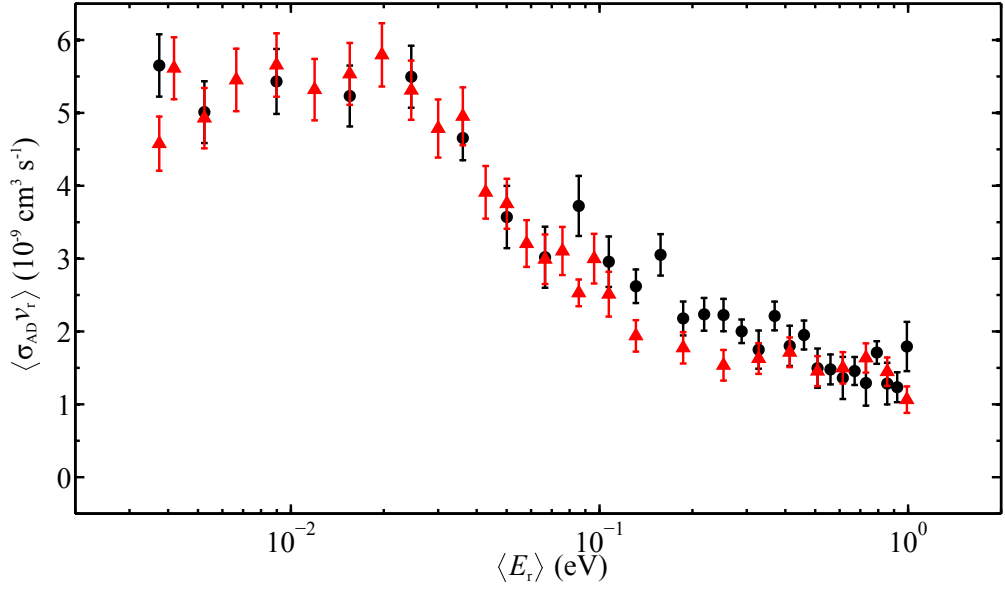}
\caption{(Color) Experimental rate coefficient $\langle \sigma_{\rm AD} v_{\rm
    r} \rangle$ as a function of the collision energy $\langle E_{\rm
    r}\rangle$.  The black circles show our new results and the red
  triangles our previous results of \cite{Holger, Bruhns2} corrected for
  the H$_2^+$ attenuation.  Although our new results extend up to
  $\langle E_{\rm r} \rangle \le 4.83$~eV, here we show only up to the
  maximum $\langle E_{\rm r} \rangle$ of our previous results for
  comparison.  The error bars show the $1\sigma$ statistical
  uncertainties.  There is an additional $\pm 12\%$ relative
  systematic error on each data point which is not shown.}
\label{fig:oldnew}
\end{figure}

Our measured rate coefficients for reaction~(\ref{eq:AD}) are plotted
in Fig.~\ref{fig:oldnew} as a function of average collision energy
$\left\langle E_{\rm{r}}\right\rangle \leq 1.0~\rm{eV}$.  The black
circles represent our new results and the red triangles our previous
work.  Both have been corrected for the attenuation of the H$_{2}^{+}$
ions.  The error bars on each data point display the 1$\sigma$
statistical uncertainty.  There is an additional $\pm 12\%$ relative
systematic error on each data point which is not shown.  The good
agreement between our new and previous results indicates that there
were no hidden systematic errors due to our previous inability to
measure and monitor $f$ and $f_{\rm nta}$ during data acquisition.

A final potential source of systematic error which we investigated was
to verify the linearity of the gas stripping method used to convert
the product H$_2$ molecules into the measured H$_2^+$ signal.  Here we
measured the AD rate coefficient as a function of helium gas cell
pressure for $(1-3)\times10^{-4}$~Torr. Table~\ref{tab:pressure} shows
the results of these AD measurements at $\langle E_{\rm r} \rangle =
16$~meV versus pressure.  Taking into account the attenuation of the
H$_{2}^{+}$ signal ions, to within the uncertainties the data show no
dependence on gas cell pressure.

Given the good agreement between our results in \cite{Holger, Bruhns2}
and our new data, we have merged them together using a
statistically-weighted averaging method.  We also included our
pressure test results in this average.  The 1$\sigma$ counting
statistics of each data point were used for the weighting.  All data
sets were also measured on the same relative energy grid.
Figures~\ref{fig:new} and \ref{fig:newlinear} show the averaged data
for $\langle E_{\rm r} \rangle \le 1$~eV plus the new data we have
collected for $1.0~{\rm eV} \le \langle E_{\rm r}\rangle \le 4.83~{\rm
  eV}$.  Also shown are the cross section calculations of
\cite{Holger, Cizek}, supplemented by our new theoretical work here,
multiplied by $v_{\rm r}$, and convolved with the experimental energy
spread.  Figure~\ref{fig:newlinear} shows the theoretical results with
and without the effects of the H$_2$ orbiting resonances included.  As
is clear from the figures, we find good agreement with theory
throughout the measured energy range.  The contribution due to
orbiting resonances of H$_2$ can also be seen in
Fig.~\ref{fig:newlinear}, as the experimental data are shifted to
slightly higher energy compared to calculations which do not include
these resonances.


\section{Discussion}
\label{sec:Discussion}

The good agreement that we find here both with our previous
results and with our updated theory strengthens our confidence that
theory and experiment have finally converged for
reaction~(\ref{eq:AD}).  Including AD via the repulsive
$^2\Sigma^+_{\rm g}$ state increases the cross section by an amount
smaller than we are currently able to measure experimentally.  The
resulting theoretical thermal rate coefficient is only 1.3\% larger
than that for only the attractive state at temperatures of 4,000~K,
3.5\% at 8,000~K and 4.4\% larger at 10,000~K.  These are
significantly smaller than the $\approx 25\%$ experimental accuracy with
which we have been able to benchmark theory.  Hence, we continue to
recommend the thermal rate coefficient of \cite{Holger} for modeling
plasma temperatures below $10^4$~K.

Additionally, our results continue to imply that the reason for the
discrepancy seen with the results of \cite{Martinez} lies in the data
of \cite{Howard} used for normalization.  This is further supported by
the theoretical AD work on hydrogen halides of \cite{Houf02a}.  They
used the same theoretical approach as we do here and found
systematically higher AD rate coefficients than the experimental work
of \cite{Howard}.  It appears to us that a re-measurement of
reaction~(\ref{eq:HCl}) using a technique different from that of
\cite{Howard} is clearly called for to resolve this dilemma.

\begin{figure}[!h]
\includegraphics[width= 1\textwidth]{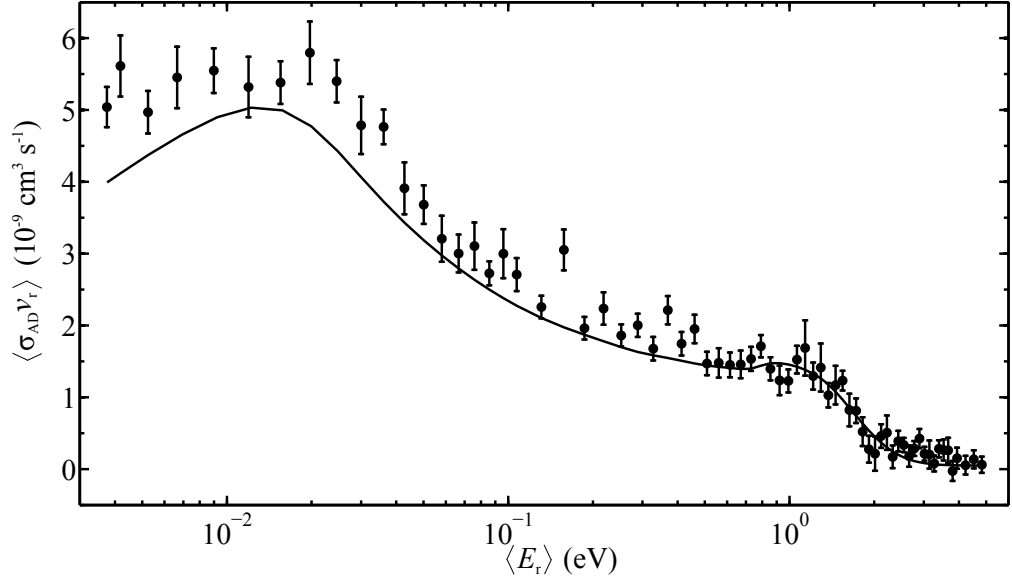}
\caption{The circles show the statistically-weighted mean of the
  experimental rate coefficients $\langle \sigma_{\rm AD} v_{\rm r}
  \rangle$ from our previous \cite{Holger, Bruhns2} and current work
  as a function of the collision energy $\langle E_{\rm r}\rangle$
  (see text).  For $\langle E_{\rm r} \rangle \geq 1.0$~eV the data
  are solely from the current measurement.  The error bars represent
  the $1\sigma$ statistical uncertainties.  The solid line is from the
  cross section calculations of \cite{Holger, Cizek}, supplemented by
  our new theoretical work here, multiplied by $v_{\rm r}$, and
  convolved with our experimental energy spread.  The effects of the
  H$_2$ orbiting resonances have been included in the calculations
  shown here.}
\label{fig:new}
\end{figure}

\begin{figure}[!h]
\includegraphics[width= 1\textwidth]{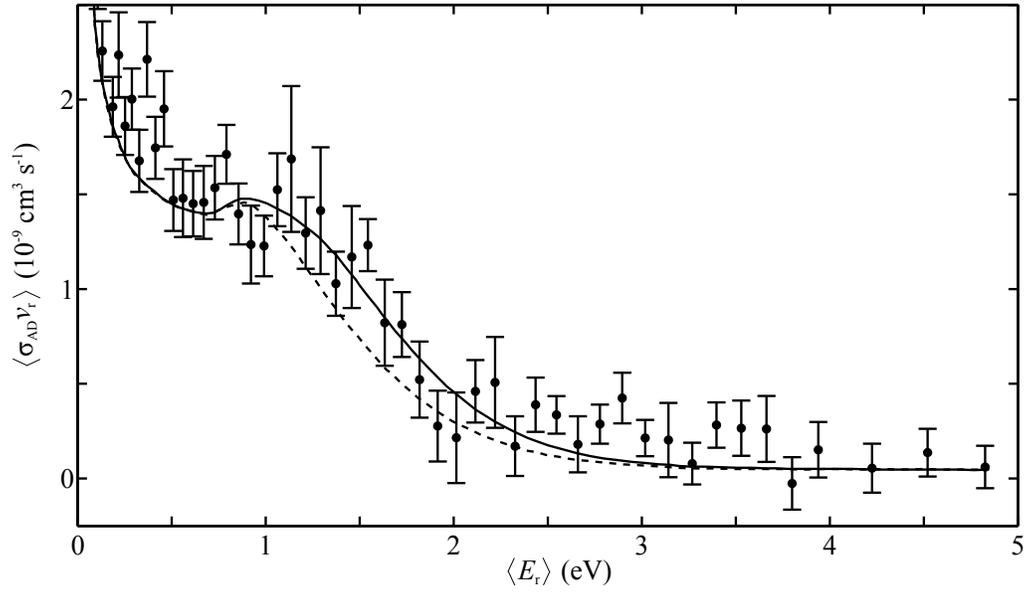}
\caption{Same as Fig.~\ref{fig:new} but on a linear scale.  The
  dotted curve shows the calculations without the effects of the
  H$_2$ orbiting resonances included.}
\label{fig:newlinear}
\end{figure}

Our results also verify the predictions of \cite{Cizek} and our new
work here that the AD cross section for reaction~(\ref{eq:AD}) should
decrease to essentially insignificant values for $E_{\rm r} \gtrsim
2$~eV, as shown in Fig.~\ref{fig:new}.  A simplified adiabatic
description of the AD reaction can provide good insight into the
physics behind this prediction.  We consider here only the
$^2\Sigma^+_{\rm u}$ symmetry.  Similar arguments can also be given
for the $^2\Sigma^+_{\rm g}$ state.

Initially the H$^-$ and H approach one another
along the attractive $^2\Sigma^+_{\rm u}$ electronic
state.  This state crosses into the
autodetachment region at $R \sim 3~a_0$.  Adiabatic theory dictates
that the system remains electronically in the ground state.  Inside
the autodetachment region the ground state is the $^1\Sigma^+_{\rm g}$
state of neutral H$_2$ plus a free electron with zero kinetic energy.
Conservation of energy requires that the final state energy equals
that initially available
\begin{equation} 
E_{v} = E_{\rm r} + D_0 -E_{\rm EA}.
\label{eq:Ev}
\end{equation}
Here $E_{v}$ is the excitation energy of the vibrational level $v$
formed in the process; $D_0$ is the 4.48~eV dissociation energy gained
by formation of H$_2$ in the $v=0$ vibrational and $J=0$ rotational
level \cite{Zhan}; and $E_{\rm EA} = 0.76$~eV is the electron affinity
required to neutralize the H$^-$ and form H \cite{Horacek}.  For
$E_{\rm r} > E_{\rm EA}$, the system lies in the dissociation
continuum ($E_{v} > D_0$), resulting in the formation of ${\rm H} +
{\rm H} + { e^-}$ and not ${\rm H_2} + { e^-}$.

In reality the AD process is not exactly adiabatic.  This is
manifested by the release of an electron with a nonzero kinetic energy
$E_{\rm e}$ and we can rewrite (\ref{eq:Ev}) as
\begin{equation} 
E_{v} + E_{\rm e} = E_{\rm r} + D_0 - E_{\rm EA}.
\label{eq:EvEe}
\end{equation}
The nonadiabatic exchange of energy between the electron and protons
is weak though; detached electrons do not have a large kinetic energy.
Our full calculations for the $^2\Sigma^+_{\rm u}$ state \cite{Cizek}
show that only a negligible amount of electrons can have energy above
$\sim 1.5$~eV.  Taking into account that the largest possible value of
$E_{\rm r}$ will occur for $E_{v} = D_0$, this leads to the prediction
that the AD process will cease for $E_{\rm r} \gtrsim E_{\rm EA} +
1.5$~eV.  For reaction~(\ref{eq:AD}), this corresponds to $E_{\rm r}
\gtrsim 2.26$~eV.  A similar argument has been suggested for the
decrease in the cross section for protonium formation in collisions of
antiprotons with hydrogen atoms (see \cite{Cohen} for a review).  Note
that we have ignored the insignificant kinetic energy of the final
H$_2$ molecule $E_{\rm H_2}$, as conservation of momentum gives
$E_{\rm H_2} = (m_{\rm e}/m_{\rm H_2})E_{\rm e} \ll E_{\rm e}$.
  
Continuing the protonium analogy, one
would expect a sharp decrease in $\sigma_{\rm AD}$ immediately
after the collisional detachment threshold at $E_{\rm r} = 0.76$~eV.
In the H$^-$ + H collision, the drop in the cross section occurs at
higher energies.  This is related to the threshold law given by
Eq.~(\ref{eq:wigner}) with $l=1$ for the dominant ungerade channel.
As a result, the coupling $V_{dk}$ vanishes for zero detached electron
energy and rises smoothly as the energy increases.  The electron
energy in Eq.~(\ref{eq:EvEe}) thus can not be exactly zero, but
remains relatively small.  The smooth decrease in $\sigma_{\rm AD}$
above 1 eV, confirmed by the present experiment, thus provides a good
test of the theoretical description of the electron release amplitude.
  
Lastly, the decrease of the AD cross section is indeed slightly
weakened by positive contributions of the $^2\Sigma_g^+$ state as
shown in Fig.~\ref{fig:sAD} and orbiting resonances as seen in
Fig.~\ref{fig:newlinear}.  However, the decreasing trend above 1eV,
controlled by $V_{dk}$, is still dominant (e.g., Fig.~\ref{fig:new}).


\section{Summary}
\label{sec:Summary}

We have modified the experimental methods used in \cite{Holger,
  Bruhns1, Bruhns2} to measure reaction (\ref{eq:AD}) up to
\textit{E}$_{\rm{r}}$~$\le$~4.83~eV. Additionally, we have performed
several modifications to better control potential systematic errors.
We find good agreement between our previous and new data sets.  To
within the experimental uncertainties, we also continue to find good
agreement with the calculations of \cite{Holger, Cizek} which have
been extended here to include contributions from the repulsive
$^2\Sigma_{\rm g}^+$ H$_2^-$ state and for the effects on the
experimental results due to orbiting resonances of H$_2$ for $E_{\rm
  r} \ge 0.76$~eV.  In particular, we confirm the predictions of
\cite{Cizek} that this reaction turns off for $E_{\rm{r}} \gtrsim
2$~eV.  Similar behavior has been predicted for the formation of
protonium from collisions of antiprotons and hydrogen atoms
\cite{Cohen}.


\begin{acknowledgments}

The authors thank M.\ Lestinsky and S.\ A.\ Marino for stimulating
discussions, and D.\ Thomas for his skilled machining work.  This work
was supported in part by NSF Grant Nos.\ CHE-0520660, AST-0606960,
AST-0807436, and AST-0905832.  H.\ Bruhns was supported in part by the
German academic exchange service DAAD.  M.\ \v{C}\'{\i}\v{z}ek and
J. Eli\'{a}\v{s}ek were supported in part by Grant No. GACR
208/10/1281 from the Czeck Republic.  X.U. acknowledges support from
the Fund for Scientific Research (FNRS)

\end{acknowledgments}


\clearpage


\clearpage


\clearpage


\clearpage


\clearpage


\clearpage


\clearpage


\clearpage


\begin{table}[!h]
\begin{center}

\caption{ Summary of systematic uncertainties at an estimated $1\sigma$ confidence level.  Uncertainties are treated as random sign errors and added in quadrature.}

\begin{tabular*}{1\textwidth}{@{\extracolsep{\fill}}      l  c   }
\hline\hline Source  & Error (\%) \\
\hline  Background subtraction & 5 \\
        Anion current  & 3 \\
        Neutral current  & 10 \\
        Beam overlap  & 3 \\
        \hline 
        Total relative errors from above & 12 \\
        Stripping cross section & 16 \\
        Effects of unknown rovibrational population  & 10 \\
        Analyzer transmittance  & 1 \\
        Grid transmittance  & 1 \\ 
        CEM detection efficiency  & 2 \\
        Overlap length  & 1 \\
        Helium gas cell column density & 7 \\
        H$_{2}^{+}$ Attenuation & 1\\
\hline  Total systematic uncertainty & 24 \\
\hline\hline
\label{tab:errors}
\end{tabular*}
\end{center}
\end{table}

\clearpage


\begin{table}[!h]
\begin{center}

\caption{ Rate coefficient results at $\langle \textit{E}_{\rm{r}} \rangle$ = 16 meV versus helium gas cell pressure. Our theoretical results are also shown for comparison. }

\begin{tabular*}{1\textwidth}{@{\extracolsep{\fill}}    c  c  c c  } \hline \hline
\multirow{2}{*}{Pressure (10$^{-4}$ Torr)  } & \multicolumn{3}{c}{ Rate Coefficient (10$^{-9}~$cm$^{3}$~s$^{-1}$) }  \\
\cline{2-4}
 &  Value &  Statistical Uncertainty &  Relative Uncertainty   \\
\hline 1.0 & 5.6  & $\pm$ 0.5 &  $\pm$ 0.7 \\
       2.0 & 5.2  & $\pm$ 0.4 &  $\pm$ 0.6 \\
       3.0 & 5.2  & $\pm$ 0.4 &  $\pm$ 0.6 \\ 
       Theory & 5.0  & -  & - \\
\hline\hline
\label{tab:pressure}
\end{tabular*}
\end{center}
\end{table}

\end{document}